\documentclass[conference]{IEEEtran}

\usepackage[OT1]{fontenc}

\usepackage{amsmath}

\usepackage[colorlinks=true, linkcolor=blue, citecolor=blue, urlcolor=blue]{hyperref}

\usepackage{graphicx}
\usepackage{booktabs}
\usepackage{multirow}
\usepackage{tabularx}
\usepackage{enumitem}
\usepackage{threeparttable}
\usepackage{balance}

\usepackage{url}

\usepackage{fancyhdr} 
\fancypagestyle{firststyle}
{
   \fancyhf{}
   \fancyfoot[C]{\scriptsize{Proceedings of the 16th Annual Conference on Privacy, Security and Trust (PST 2018), Belfast, IEEE, pp.~1--5, \url{https://doi.org/10.1109/PST.2018.8514163}.}}
}

\begin{document}

\title{Crossing Cross-Domain Paths in the Current Web}
\author{
\IEEEauthorblockN{Jukka Ruohonen}
\IEEEauthorblockA{University of Turku, Finland \\
Email: juanruo@utu.fi}
\and
\IEEEauthorblockN{Joonas Salovaara}
\IEEEauthorblockA{University of Turku, Finland \\
Email: joonas.salovaara@gmail.com}
\and
\IEEEauthorblockN{Ville Lepp\"anen}
\IEEEauthorblockA{University of Turku, Finland \\
Email: ville.leppanen@utu.fi}
}

\maketitle

\begin{abstract}
The loading of resources from third-parties has evoked new security and privacy concerns about the current world wide web. Building on the concepts of forced and implicit trust, this paper examines cross-domain transmission control protocol (TCP) connections that are initiated to domains other than the domain queried with a web browser. The dataset covers nearly ten thousand domains and over three hundred thousand TCP connections initiated by querying popular Finnish websites and globally popular sites. According to the results, (i)~cross-domain connections are extremely common in the current Web. (ii)~Most of these transmit encrypted content, although mixed content delivery is relatively common; many of the cross-domain connections deliver unencrypted content at the same time. (iii)~Many of the cross-domain connections are initiated to known web advertisement domains, but a much larger share traces to social media platforms and cloud infrastructures. Finally, (iv) the results differ slightly between the Finnish web sites sampled and the globally popular sites. With these results, the paper contributes to the ongoing work for better understanding cross-domain connections and dependencies in the world wide web.
\end{abstract}


\section{Introduction}

\thispagestyle{firststyle} 

In February 2018 a Belgian court ordered the social media giant Facebook to stop collecting data about Belgian citizens via third-party sites, cookies, and invisible pixels~\cite{Guardian18}. These tracking techniques are all well-know. When considering the tracking of a user from a website to another, privacy issues are present already with the \texttt{Referer} hypertext transfer protocol (HTTP) header field often sent to a landing page from a current page. Cookies and pixels help. This said, it is JavaScript that supply most of the dynamic functionality of contemporary websites---and likely delivers most of the privacy violations and security issues~\text{\cite{Malandrino13, Kayes17}}. Potential security issues are evident already because external JavaScript has full privileges within the requesting website with few exceptions~\cite{Ruohonen18IFIPSEC}. It also the JavaScript content loaded from Facebook's servers that raises privacy concerns when visiting conventional websites that incorporate innocent-looking web elements such as Facebook-buttons~\cite{SomeBielova17, Simonds12}. These elements can be linked to two fundamental theoretical concepts.

The first is the concept of \textit{forced trust}, which refers to situations in which users have no choice or opportunity to affect information systems---including even the choice to (not) use a given system~\cite{Hakkala17}. Given that Facebook-buttons are commonly placed even on the websites of law enforcement agencies and online banking sites, a user has practically no choice but to accept the possibility that traces are left to social media platforms upon paying bills or contacting a police officer. The second concept is \textit{implicit trust}, which refers to situations in which websites' operators are implicitly trusting unknown third-parties by explicitly trusting known parties. In particular, the claim is that ``websites operators no longer know who they are trusting because external services load \textit{implicitly trusted} content from third parties that are unknown to the main site operator''~\cite{Kumar17}. Although the claim may be provocative to some extent, the theoretical implications are vicious.

\begin{figure}[th!b]
\centering
\includegraphics[width=6.5cm, height=4cm]{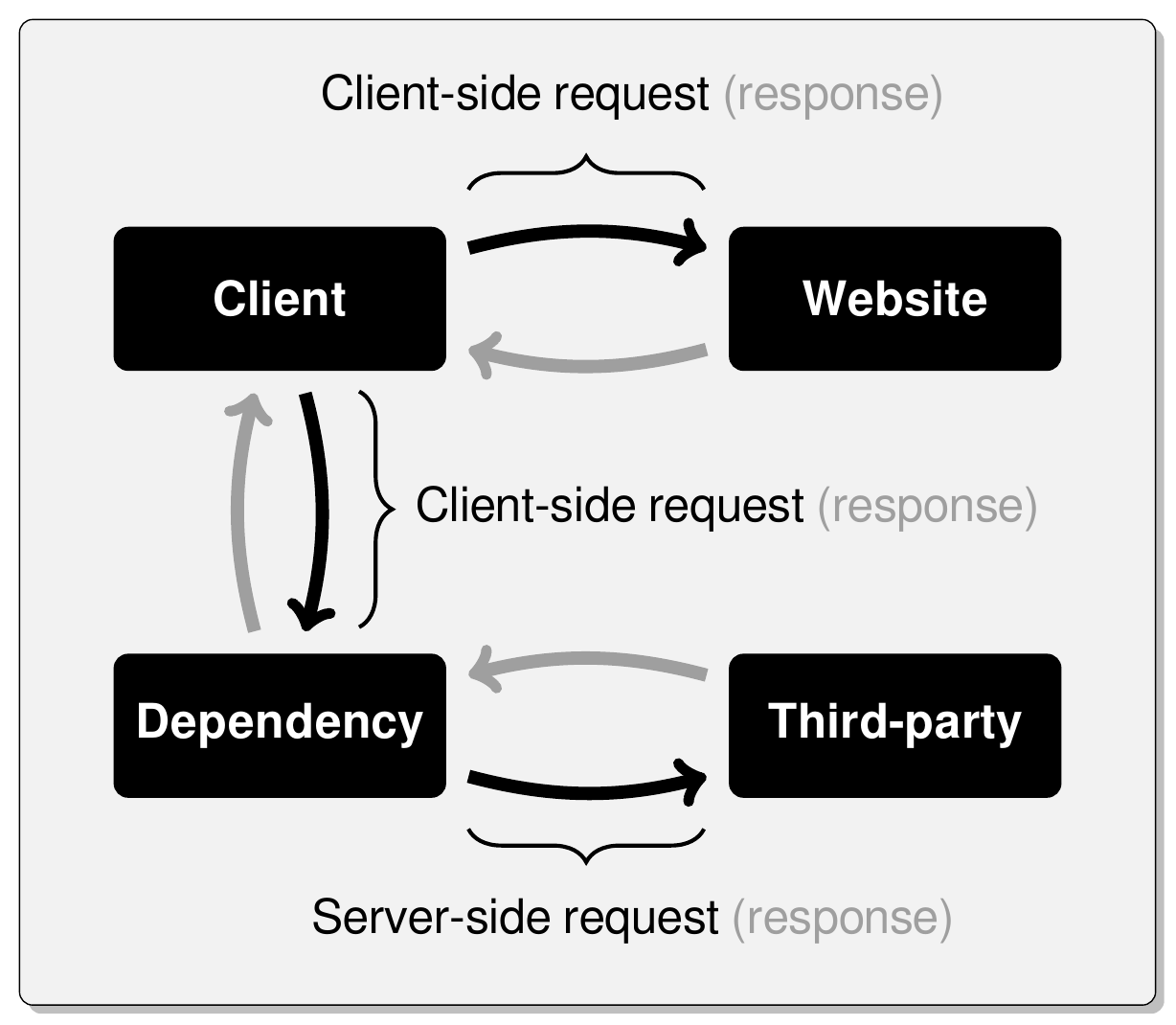}
\caption{An Example of Implicit and Forced Trust}
\label{fig: example}
\end{figure}

When the forced and implicit trust are combined, it is not difficult to draw intriguing theoretical scenarios about the implications. As an example: a user (citizen) is forced to trust an online information system of a law enforcement agency, the agency's system is explicitly trusting content loaded from a known party, and the known party is loading additional content from a party unknown to the agency and the user---hence, the user is forced to trust an unknown party because the agency's system is implicitly trusting the unknown party. The scheme in Fig.~\ref{fig: example} illustrates these two theoretical concepts further. In this simple scheme, a user visits a website that contains a known dependency to another website. Once the user's client requests the dependency from the known website, this website makes a further server-side request \cite{Pellegrino16} to a third-party whose response is subsequently delivered to the client. Thus, neither the client nor the operator of the website necessarily know the third-party, although it is implicitly trusted by the operator.

Motivated by analogous theoretical scenarios and more traditional privacy questions, there has recently been an increasing interest to empirically examine dependencies between websites   \text{\cite{SomeBielova17, Kumar17, Metwalley16}}. While new questions related to third-parties have emerged with social media \cite{Kayes17}, the empirical research in this domain has its roots in the long-standing questions related to web advertisements, user tracking, and the never-ending ``ad-blocking wars''~\text{\cite{Achara16, Gervais17, Wills16}}. This is also the domain to which this paper contributes with the focus on TCP connections and Finnish websites. Although this low-level focus allows to only observe domain names, Internet protocol addresses, and ports, it should be kept in mind that much more information is leaked with cross-domain connections. When the leaks are combined with geolocation tracking and browser fingerprinting, identifying unique users is likely easy particularly in small countries such as Finland. Given these remarks, four research questions (RQs) are examined:

\begin{enumerate}[label={RQ$_{\arabic{enumi}}$}]
\itemsep0.2em
\item{\textit{How common are cross-domain TCP connections when paying visits to popular world wide websites?}}\label{rq: cross-domain}
\item{\textit{How common is mixed-content web delivery that uses both HTTP and HTTPS (a.k.a.~HTTP over SSL/TLS)?}}\label{rq: mixed}
\item{\textit{When paying visits to popular world wide websites, how common are involuntary cross-domain connections to social media platforms and advertisement domains?}}\label{rq: some-ads}
\item{\textit{Do the answers to \ref{rq: cross-domain}, \ref{rq: mixed}, and \ref{rq: some-ads} differ between popular Finnish websites and globally popular sites?}}\label{rq: finnish}
\end{enumerate}

To answer to these four questions, the paper proceeds by introducing the dataset in Section~\ref{section: data}. Results are presented in Section~\ref{section: results}, and discussion follows in the final Section~\ref{section: discussion}.

\section{Data}\label{section: data}

In what follows, the empirical sample is elaborated by discussing the data collection routine and the domains queried.

\subsection{Queries and Captures}

\begin{figure}[th!b]
\centering
\includegraphics[width=7cm, height=3.6cm]{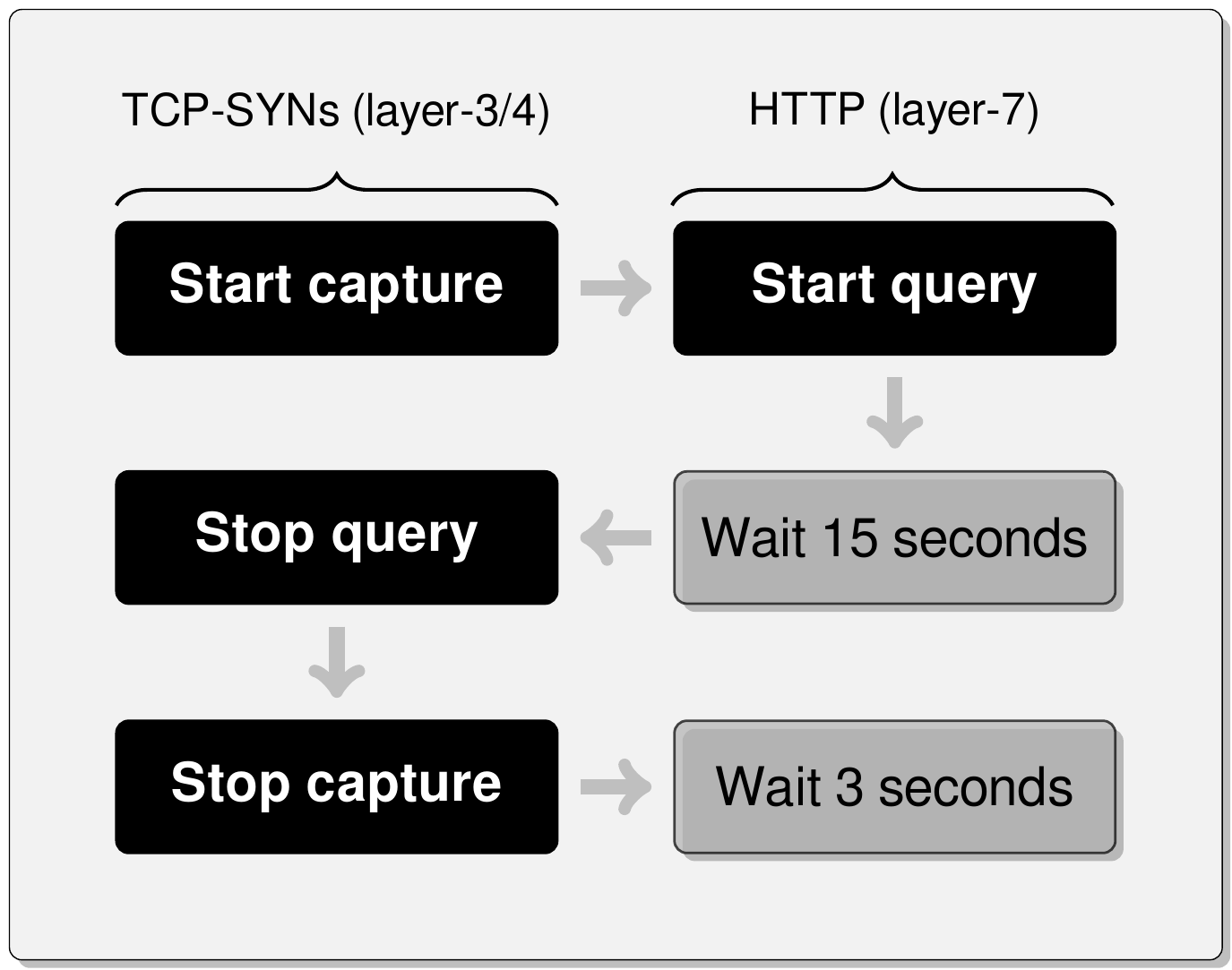}
\caption{A Simple Routine for Capturing \& Querying (C\&Q)}
\label{fig: queries}
\end{figure}

There are two common approaches for collecting and quantifying cross-domain information. The first \textit{active} approach uses headless browsers, browser plugins, or proxies alongside JavaScript instrumentation, hypertext parsing, and related means \cite{SomeBielova17, Kumar17, Gervais17, ChenNikiforakis15}. The second \textit{passive} approach relies on HTTP header information and lower-level network traces~\cite{Metwalley16, Pujol15, Metwalley15}. The passive approach is used in this paper: before launching a query with a web browser, the classical \texttt{tcpdump} program~\cite{tcpdump} is started for capturing the initial synchronization (SYN) phase in the three-way handshake used to establish new TCP connections. All other packets are disregarded. While the C\&Q routine illustrated in Fig.~\ref{fig: queries} is simple, it is idiot-proof in the sense that all new TCP connections are guaranteed to be captured, which would not necessarily be the case with active parsing-based approaches.

The querying was done with Firefox $58.0.1$. The browser's default settings were used with the exception of configurations needed to exclude all connections initiated by default. These include telemetry pings to Mozilla's servers and different connections related to safe browsing and malicious websites. After each query, all Firefox-specific data was flushed, sans the noted configurations. Due to the dynamic nature of the contemporary Web, each query was followed by a fifteen second delay. This delay is sufficient to ensure that full contents were loaded, including dynamic content generated through JavaScript. It should be further noted that name resolution was also left for \texttt{tcpdump}, which in this case uses the live domain name system (DNS) through a resolver provided by a local Internet service provider. Before running the C\&Q routine,  the Linux operating system was also configured to limit the retries for TCP-SYN packets to one. Even with this precaution, a small possibility of double-counting must be acknowledged.

\subsection{Samples}

Two domain name lists are used for the data collection with the C\&Q routine. The first list contains the top-10000 popular unique second-level domains extracted from a larger list made available by Cisco~\cite{Cisco18}. The second list is based on previous research \cite{Ruohonen17EISIC} on Finnish websites; the list contains a little over two hundred popular websites that serve content primarily in the Finnish language. Unlike Cisco's list based on the volume of DNS traffic in the global Internet, the Finnish list is compiled by a market research company~\cite{TNS17} for the explicit goal of ranking Finnish websites. The ranking is done with cookie-tracking, traffic analysis, surveys, and related means.

\begin{table}[th!b]
\centering
\caption{Sample Characteristics}
\label{tab: sample}
\begin{small}
\renewcommand*{\arraystretch}{1.2}
\begin{tabular}{lrrr}
\toprule
& \multicolumn{3}{c}{Sample} \\
\cmidrule{2-4}
& Finnish && Cisco \\
\hline
Domains (websites) & $210$ && $9216$ \\
Connections (TCP-SYNs) & $18634$ && $307516$ \\
Connections to ports $\not\in\lbrace 80, 443 \rbrace$ & $0$ && $112$ \\
\bottomrule
\end{tabular}
\end{small}
\end{table}

Both lists were pre-processed by transforming the domain names into second-level domains. This commonly used manipulation (e.g., \cite{Wills16}) is both a reasonable simplification and a necessity due to the DNS-based rankings in the Cisco sample. For instance, \texttt{google.com} takes the fourth place and \texttt{www.google.com} the fifth place in the list, but for the purposes of this paper, the two domains are equal. Given this manipulation, over nine thousand successful queries were made to the (second-level) domains in the Cisco sample. As can be concluded already from the numbers in Table~\ref{tab: sample}, these queries required a very large amount of new TCP connections.

\section{Results}\label{section: results}

The empirical results are presented consecutively by considering the answers to the three research questions \ref{rq: cross-domain}, \ref{rq: mixed}, and  \ref{rq: some-ads}. The answer to \ref{rq: finnish} is contemplated along the way.

\subsection{Cross-Domain Connections (\ref{rq: cross-domain})}

If one term should be picked for describing the infrastructure side of today's Web, cross-domain dependencies would be a good choice. According to recent estimates based on the Alexa top-million sites, over 90\% of popular sites have dependencies to external domains, the median of external resources loaded is 73, about two thirds of all resources are loaded from third-parties, and perhaps most strikingly, these cross-domain dependencies largely trace only to four technology companies~\cite{Kumar17}. Three of these companies---Google, Facebook, and Twitter---are also attracting a large share of the speculative investments for web technologies \cite{Newman16}. These general insights about cross-domain dependencies and their concentration provide a good way to start the empirical analysis.

\begin{figure}[th!b]
\centering
\includegraphics[width=\linewidth, height=4cm]{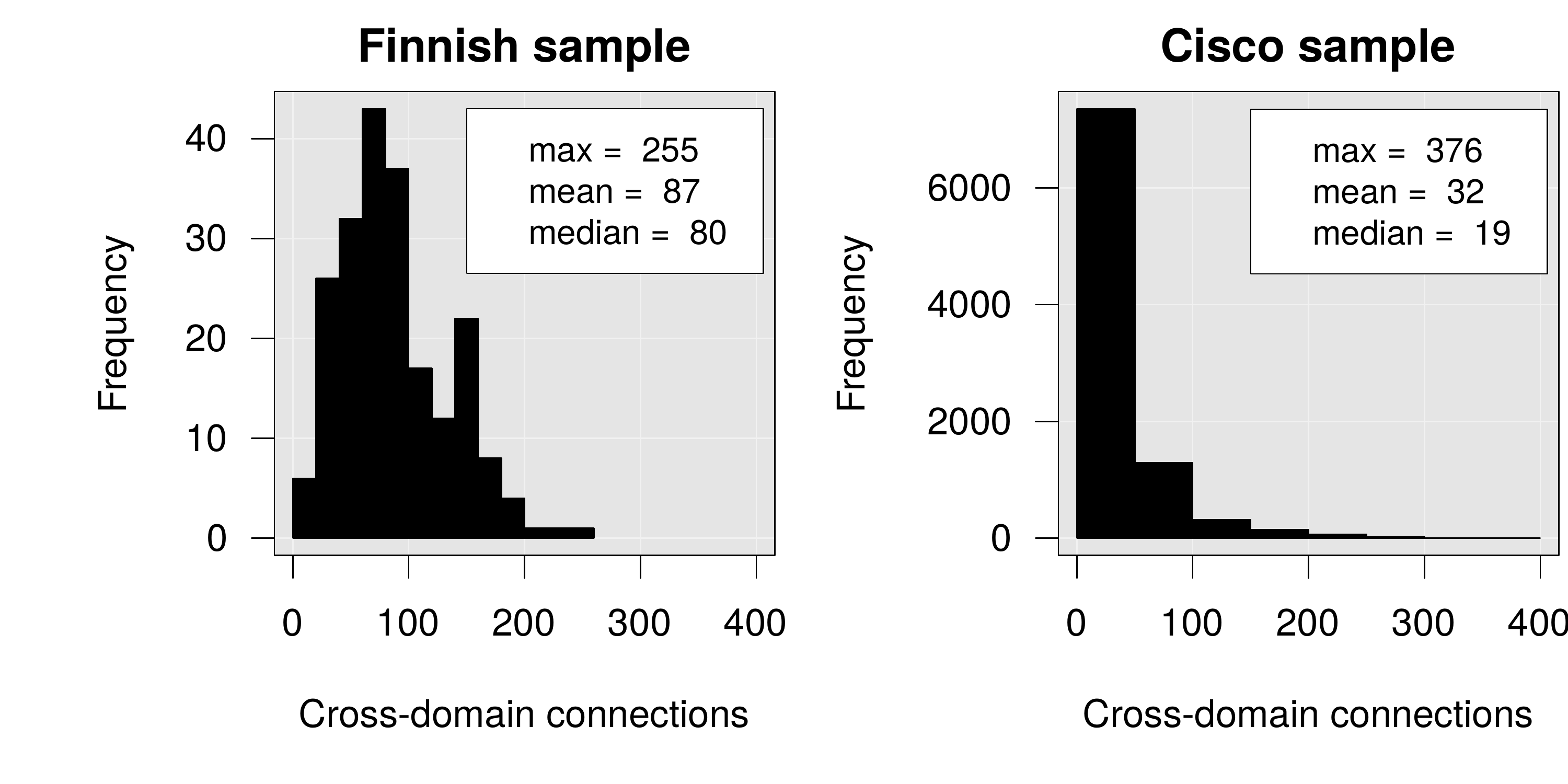}
\caption{Cross-Domain TCP Connections Initiated}
\label{fig: crosscon}
\end{figure}

In theory, the same-origin policy \cite{RFC6454} could be used for quantifying cross-domain connections. However, as the data captured only contains traces about TCP connections initiated, a more relaxed definition is used. Thus, for each domain queried, a cross-domain connection is defined simply as a TCP packet with a SYN flag sent to a domain other than the domain queried. As the domains queried refer to second-level domain names, also the comparison is done based on second-level domain names. It should be further noted that the definition is rather lax because many popular domains are aliased via DNS to content delivery networks (CDNs) and cloud services. For instance, at the time of writing, the Finnish domain \texttt{yle.fi} had eight IPv4 addresses, all of which but one pointed to subdomains of Amazon's \texttt{cloudfront.net} according to reverse (PTR) DNS lookups. Therefore, as many as 93\% of the TCP connections initiated when querying \texttt{yle.fi} are classified as being cross-domain, although, in a sense, the connections to \texttt{cloudfront.net} might not be counted as cross-domain connections because the primary delivery is done through Amazon. While keeping this point in mind, the results can be summarized with the two histograms shown in Fig.~\ref{fig: crosscon}.

The amount of \textit{new} cross-domain connections initiated is substantial: the median is $19$ in the Cisco sample and as much as $80$ in the Finnish sample. In both samples, some of the domains queried entailed even up to $250$ cross-domain TCP connections. These numbers are substantial because the so-called HTTP/1.1 keep-alive option allows to load multiple resources without initiating new TCP connections. The apparent difference between the samples is also interesting. One explanation relates to the unequal sample sizes: if more Finnish domains would be queried, the results might eventually converge toward a long-tailed distribution seen in the Cisco sample. Another, competing explanation relates to the fact that most of the third-party web resources are hosted from the United States \cite{Gervais17}. Furthermore: unlike many of the domains in the Cisco sample, the Finnish domains do not have their own globally operating CDNs or cloud infrastructures, which increases the amount of cross-domain connections observed in the Finnish sample.

\subsection{Mixed Delivery (\ref{rq: mixed})}

\begin{figure}[th!b]
\centering
\includegraphics[width=\linewidth, height=8cm]{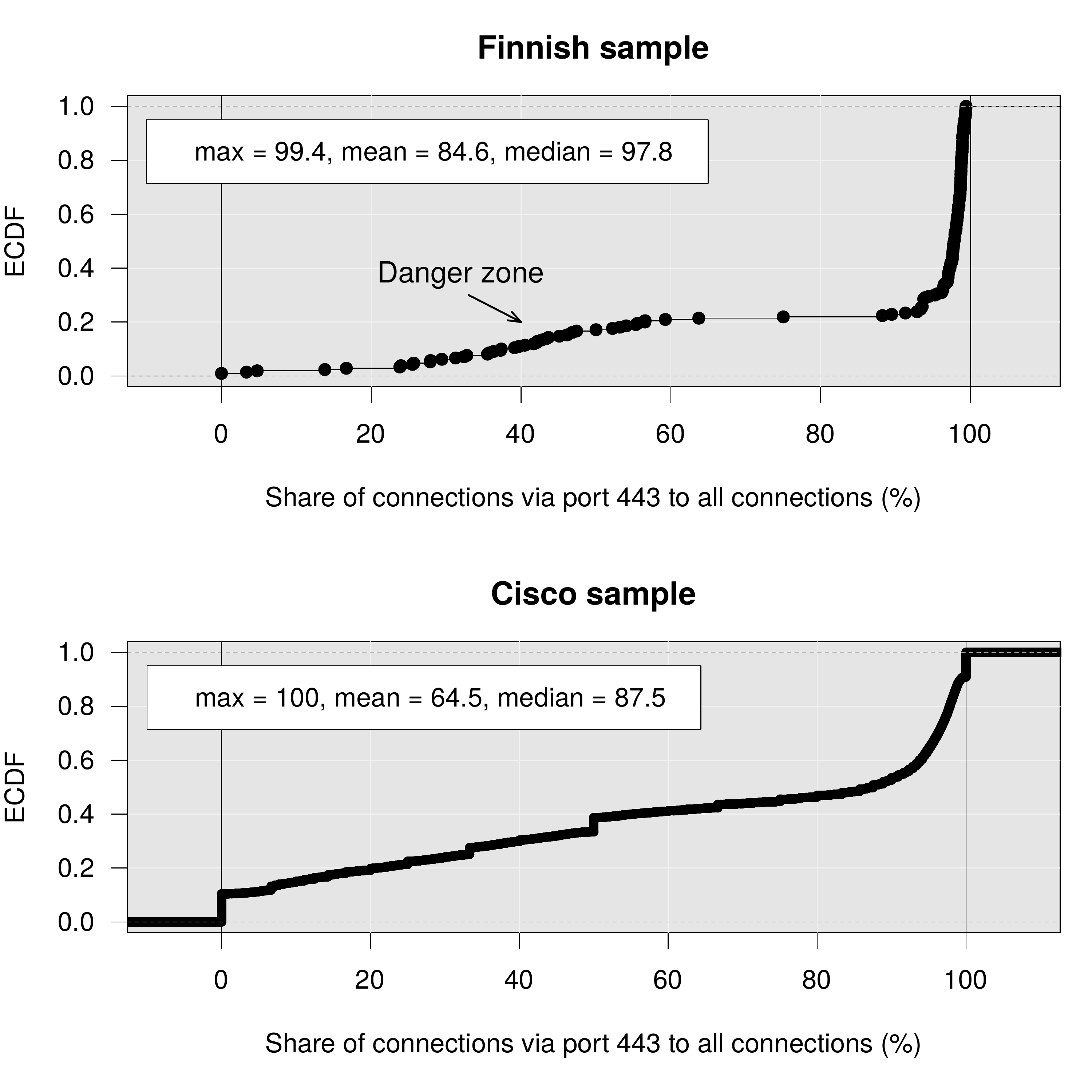}
\caption{Mixed Connections Initiated}
\label{fig: mixed}
\end{figure}

To summarize the extent of mixed content delivery, the amount of new HTTPS connections is approximated by counting the number of connections initiated through the default port $443$. The per-domain percentage share of these connections to all TCP-SYNs sent is visualized in Fig.~\ref{fig: mixed} by using the empirical cumulative distribution function (ECDF).

Both samples indicate that HTTPS is widely used in the current Web: the medians are $98\%$ and $88\%$ in the Finnish and Cisco samples. Thus, on average, the encryption situation is better among popular Finnish domains compared to the globally popular domains. Interestingly, however, none of the Finnish domains reach the maximum of 100\%, meaning that at least one HTTP connection was required for a redirection from HTTP to HTTPS. In contrast, all connections were initiated through HTTPS for many domains in the Cisco sample. A~possible explanation traces to a list loaded by Firefox \cite{Keeler12} to counter a bootstrap problem \cite{ChenNikiforakis15} by ensuring that also initial connections are encrypted and comply with the HTTP strict transport security (HSTS) standard \cite{RFC6797}. In essence, HSTS can be used to instruct browsers to transform insecure (\texttt{http://...}) links to secure (\texttt{https://...}) ones. According to the results shown in Fig.~\ref{fig: mixed}, a wider server-side adoption of HSTS would be welcome on the client-side.

That is, both samples contain many domains with a large share of mixed content delivery. Although the data captured does not allow to speculate about the content transmitted, most of these deliveries are presumably images, cascading style sheets, and related elements loaded particularly from CDNs. Security risks are a lesser concern for such content. But when also JavaScript is loaded either from CDNs or other third-party domains via plain HTTP, there is a risk of a man-in-the-middle attack scenario~\cite{Kumar17}. Cookie stealing, request forgery, and sensitive information leaks are also possible~\cite{ChenNikiforakis15}. Given that bugs occur during development and mistakes are made during maintenance, the risks may be real for the few observed websites that try to tightrope between HTTP and HTTPS instead of delivering everything through the latter protocol.

\subsection{Advertisers and Social Media (\ref{rq: some-ads})}

For providing a tentative answer to \ref{rq: some-ads}, the cross-domain connections initiated are compared against two so-called ad-blocking lists \cite{AdlistA18, AdlistB18}, and two lists on the domains owned and used by Facebook and Google~\cite{Blocklists17}. It should be remarked that particularly the ad-blocking lists contain many problems, including the manual maintenance~\cite{Achara16} and the variance between lists~\cite{Gervais17, Wills16}. The two lists used should still provide a reasonable approximation on the prevalence of advertisement servers among the servers to which new cross-domain TCP connections were initiated. As has been common \cite{Metwalley15}, the matching is again done based on second-level domain names.

The median of cross-domain connections to ad-servers is only two in the Finnish sample and zero in the Cisco sample. However, about 54\% of the domains in the Finnish sample and 19\% of the globally popular domains entailed at least one connection to a domain serving advertisements. Particularly the high share in the Finnish sample is noteworthy. The explanation presumably relates to the fact that the most popular Finnish sites are owned by media companies whose revenues are partially dependent on the web advertisements served.

\begin{figure}[th!b]
\centering
\includegraphics[width=\linewidth, height=4cm]{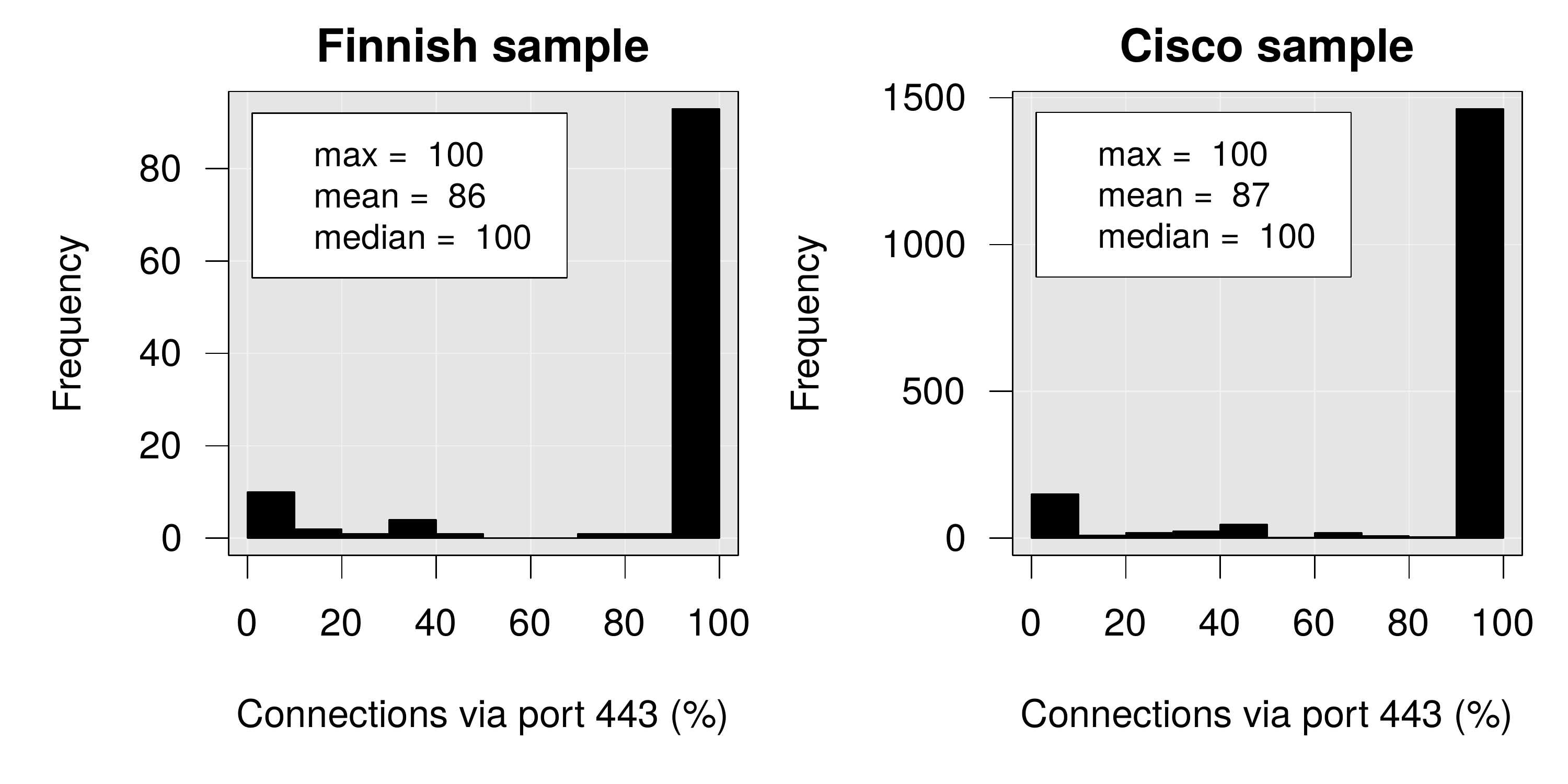}
\caption{Mixed Connections to Advertisement Domains}
\label{fig: ads https}
\end{figure}

\begin{figure}[th!b]
\centering
\includegraphics[width=\linewidth, height=4.5cm]{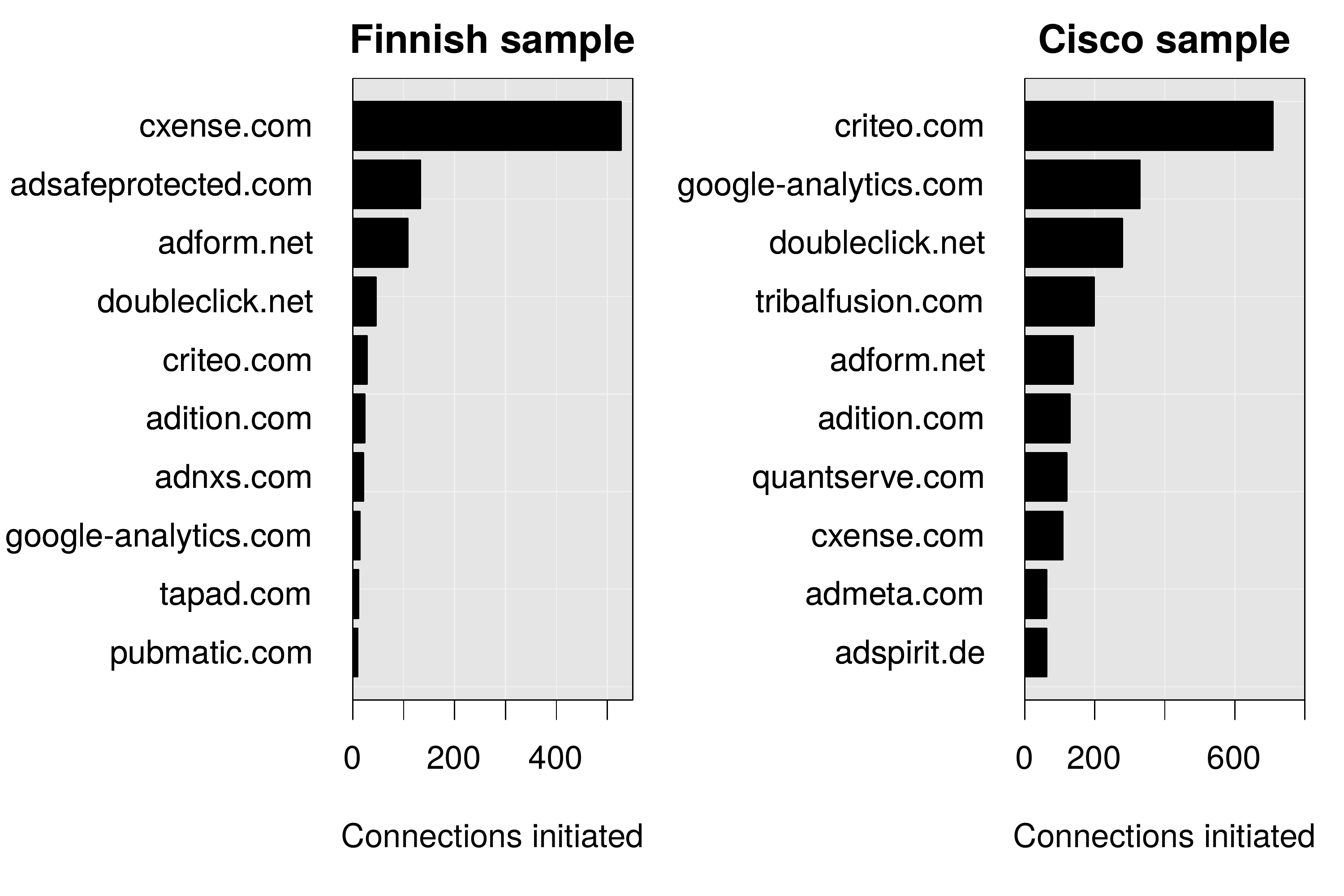}
\caption{Cross-Domain Connections to Top-$10$ Advertisement Domains}
\label{fig: ads}
\end{figure}

\begin{figure}[th!b]
\centering
\includegraphics[width=\linewidth, height=7cm]{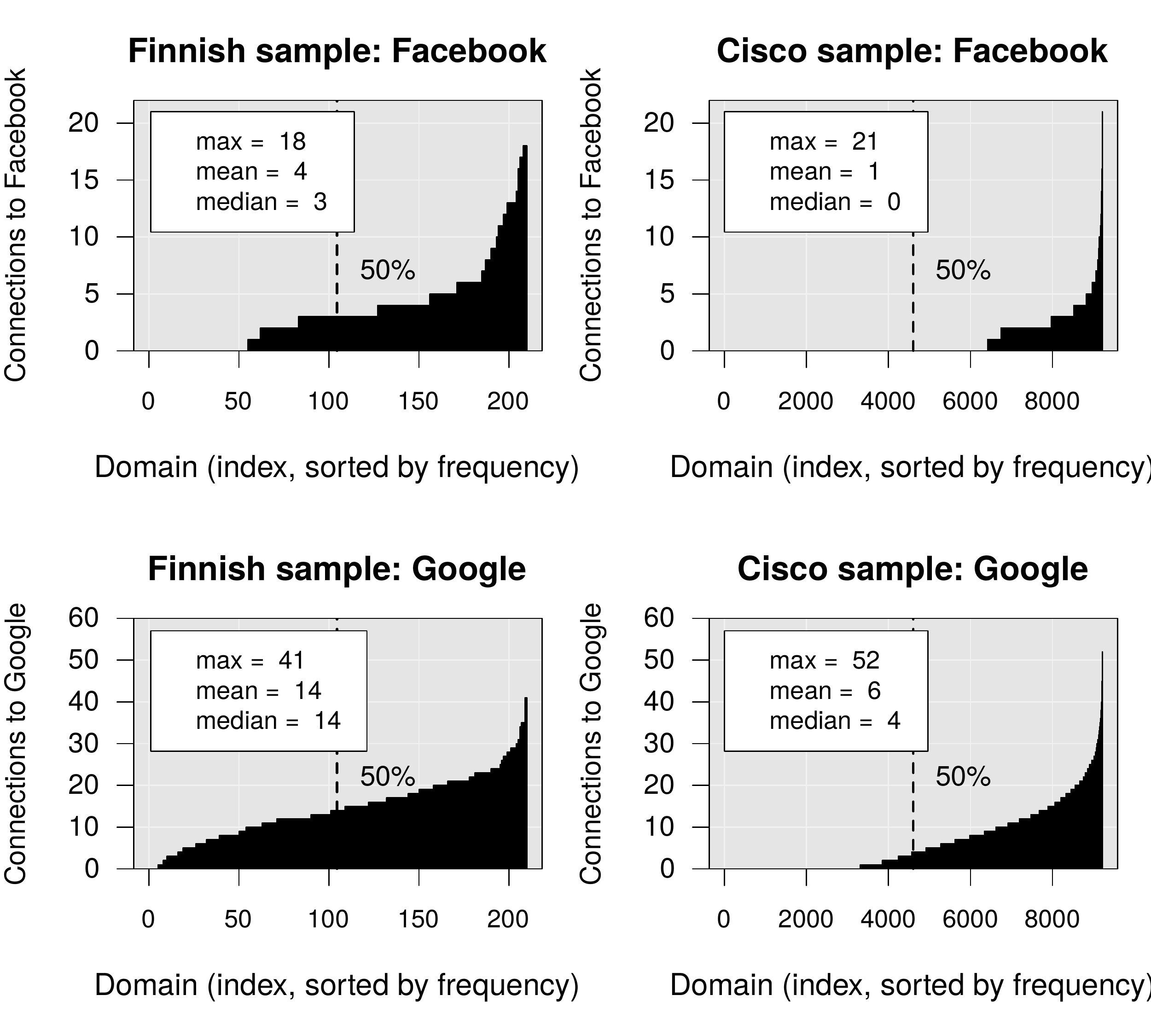}
\caption{Cross-Domain TCP Connections Initiated to Facebook and Google}
\label{fig: fb google}
\end{figure}

Another point is that almost all of the connections to advertisement domains occurred through HTTPS (see Fig.~\ref{fig: ads https}). While these shares are much higher than what has been observed in the past, the conclusion remains identical: also advertisers follow the general HTTPS adoption trends~\text{\cite{Metwalley16, Metwalley15}}. Furthermore, the most common advertisement domains diverge between the two samples (see Fig.~\ref{fig: ads}). While this observation is expected due to geographic market structures in the web advertisement industry, the results  differ also from a cookie sample recently collected from the same Finnish domains~\cite{Ruohonen17EISIC}. The reason may relate to so-called ad-exchanges that act as brokers between website owners and advertisement companies. Consequently, the amount of advertisements displayed on a website may remain relatively constant in the short-run, but the actual domains from which these are served may vary.

As can be seen from the final Fig.~\ref{fig: fb google}, it is not necessarily the web advertisements that raise the largest privacy concerns, however. Clearly over a half of the domains in the Finnish sample entailed at least one TCP connection to Facebook's and Google's servers. While there are again differences between the two samples, the shape of the four distributions is similar; at the tails, there are domains that initiate even over forty cross-domain connections toward Google. The rates are lower but still substantial for Facebook. Insofar as privacy is a concern for users, it is not necessarily \texttt{google-analytics.com} and \texttt{doubleclick.net} that users should worry about. Arguably, it is rather the loading of fonts, images, scripts, and other web resources from cloud services, CDNs, and other middlemen of the current Web. 

\section{Discussion}\label{section: discussion}

This paper examined cross-domain TCP connections initiated when visiting popular websites. To answer (A) to the four research questions, the results can be summarized as follows:
\begin{enumerate}[label={A$_{\arabic{enumi}}$}]
\itemsep0.2em
\item{\textit{The amount of cross-domain connections is substantial: most websites require tens of connections to other domains than the domains on which the sites are located.}}\label{a: cross-domain}
\item{\textit{Mixed content delivery occurs, although most of the cross-domain connections are done through HTTPS.}}\label{a: mixed}
\item{\textit{Many cross-domain connections are initiated to known web advertisement servers, but a much larger amount traces to social media platforms such as Facebook and cloud infrastructures such as the one provided by
Google.}}\label{a: some-ads}
\item{\textit{The answers \ref{a: cross-domain}, \ref{a: mixed}, and \ref{a: some-ads} differ somewhat between popular Finnish websites and globally popular sites.}}
\end{enumerate}

Particularly the Finnish sample aligns well with previous observations regarding Google's almost perfect penetration across the current Web \cite{Metwalley16}. When visiting a popular Finnish website, the probability is very close to one that some traces are left to Google's servers. Even when a user is not willing to voluntarily participate in the Facebook's \textit{Panopticon} \cite{Fuchs11}, the probability is nearly $0.8$ for involuntary traces to appear also in Facebook's servers upon visiting a popular Finnish website.

As social media platforms are under increasing scrutiny particularly in Europe, further empirical research is required not only for examining privacy questions on the client-side, but also for extending the questions toward regulatory aspects and the concept of forced trust. For further research on the side of implicit trust and security, it seems prolific to turn the attention toward the actual source code of the JavaScript possibly loaded from unknown third-parties via known parties.

\bibliographystyle{IEEEtran}


\end{document}